\documentclass[sigconf]{acmart}
\renewcommand
\footnotetextcopyrightpermission[1]{}
\usepackage{hhline}
\AtBeginDocument{%
  \providecommand\BibTeX{{%
    \normalfont B\kern-0.5em{\scshape i\kern-0.25em b}\kern-0.8em\TeX}}}

\copyrightyear{}

\acmConference[]{}{}{}

\begin{document}


%
\title{A Large-Scale Rich Context Query and Recommendation Dataset in Online Knowledge-Sharing  }



\author{Bin Hao$^1$, Min Zhang$^1$*, Weizhi Ma $^1$, Shaoyun Shi$^1$, Xinxing Yu$^2$, Houzhi Shan$^2$, Yiqun Liu$^1$  and Shaoping Ma$^1$}
\affiliation{$^1$Department of Computer Science and Technology, Institute for Artificial Intelligence,\\
Beijing National Research Center for Information Science and Technology, Tsinghua University, Beijing, China\\
$^2$Zhihu, Beijing, China\\
haob15@mails.tsinghua.edu.cn, z-m@tsinghua.edu.cn
}
\renewcommand{\shortauthors}{Bin Hao, et al.}
\begin{abstract}
Data plays a vital role in machine learning studies.  In the research of recommendation, both user behaviors and side information are helpful to model users. So, large-scale real scenario datasets with abundant user behaviors will contribute a lot.
However, it is not easy to get such datasets as most of them are only hold and protected by companies.
In this paper, a new large-scale dataset collected from a knowledge-sharing platform is presented, which is composed of around 100M interactions collected within 10 days, 798K users, 165K questions, 554K answers, 240K authors, 70K topics, and more than 501K user query keywords. There are also descriptions of users, answers, questions, authors, and topics, which are anonymous. Note that each user's latest query keywords have not been included in previous open datasets, which reveal users' explicit information needs.

We characterize the dataset and demonstrate its potential applications for recommendation study.  Multiple experiments show the dataset can be used to evaluate algorithms in general top-N recommendation, sequential recommendation, and context-aware recommendation. This dataset can also be used to integrate search and recommendation and recommendation with negative feedback. Besides, tasks beyond recommendation, such as user gender prediction, most valuable answerer identification, and high-quality answer recognition, can also use this dataset. To the best of our knowledge, this is the largest real-world interaction dataset for personalized recommendation.
\end{abstract}






\maketitle
\let\thefootnote\relax\footnote{* Corresponding Author}
\let\thefootnote\relax\footnote{This work is supported by the National Key Research and Development Program of China (2018YFC0831900), Natural Science Foundation of China (Grant No. 62002191, 61672311, 61532011) and Tsinghua University Guoqiang Research Institute. This project is also funded by China Postdoctoral Science Foundation (2020M670339) and Dr. Weizhi Ma has been supported by Shuimu Tsinghua Scholar Program.}
\section{INTRODUCTION}
In recent years, with the development of the Internet industry, there is a huge amount of information generated at all times, and information overload has also arisen. How to help users find information that meets their needs is an important issue, and the personalized recommendation system \cite{sharma2013survey} is considered as an important way to solve this problem. At present, personalized recommendation technology is playing a more and more important role in many scenarios, such as shopping, video, reading, social networking, etc. 

However, it is hard for researchers to get open large-scale real scenario datasets with abundant user interactions for the recommendation study, as most of them are only hold and protected by big companies. Most of the existing open large-scale datasets, such as Amazon dataset\footnote{http://jmcauley.ucsd.edu/data/amazon/}, MovieLens dataset\footnote{https://grouplens.org/datasets/movielens/}, are in e-commerce or movie scenario, and lack user interaction logs with the recommendation systems.

This paper presents a new dataset, named ZhihuRec, to fill this gap, which is collected from Zhihu\footnote{http://www.zhihu.com/}, a socialized knowledge-sharing community. Unlike previous scenarios, the socialized knowledge-sharing community allows users to share high-quality knowledge in the community by asking and answering questions, which has rich textual information and a genuine semantic relationship between questions and answers. Users are shown with a recommendation Q$\&$A list when entering this system, and they can search Q$\&$As with free queries, so both recommendation and query behaviors exist in this scenario.

ZhihuRec is a large dataset with detailed feature information in a socialized Q$\&$A scenario. It retains complete user interactions (e.g., click, skip, query, etc.), timing, and content information. This allows researchers to verify the performance of different types of recommendation algorithms, such as collaborative filtering, content-based recommendation, sequential-based recommendation, knowledge-enhanced recommendation, and hybrid recommendation.

Besides, due to the abundant information in ZhihuRec, it can be applied to not only recommendation studies but also user modeling (e.g., gender prediction, user interest prediction), the combination of search \& recommender system, and other interesting topics.

There are mainly three advantages of the ZhihuRec dataset:

(1) To the best of our knowledge, ZhihuRec is the largest public recommendation dataset, containing various user interactions collected from an online knowledge-sharing community called Zhihu. The dataset is publicly available and can be downloaded from https://github.com/THUIR/ZhihuRec-Dataset.

(2) ZhihuRec dataset provides abundant content information, including questions, answers, profiles, topics. Especially, users' query logs are also revealed, which has not been contained before.

(3) Besides recommendation studies such as top-N recommendation, context-aware recommendation, ZhihuRec can be used in various research areas, such as user modeling and integrating of search and recommendation study.


\begin{table*}[]
\centering
\caption{The differences between ZhihuRec and some popular recommendation datasets. $\circ$ denotes that the ZhihuRec dataset contains the \#favorite of answers}
\begin{tabular}{|c|c|c|c|c|c|c|c|c|}
\hline
\textbf{Dataset}               & Textual Content    & User Profile & Item Attributes & Timestamps & Impression & Click   & Search query & rating/favourite\\ \hline
\textbf{Movielens}             &         & $\surd$ & $\surd$   & $\surd$               &  &    &        &   $\surd$\\ \hline
\textbf{Amazon}                & $\surd$ &         & $\surd$   & $\surd$               &  &    &        &   $\surd$\\ \hline
\textbf{Yelp}                  & $\surd$ & $\surd$ & $\surd$   & $\surd$               &  &    &        &   $\surd$ \\ \hline
\textbf{Xing} &         & $\surd$ & $\surd$   & $\surd$   & $\surd$    & $\surd$       &   &      \\ \hline
\textbf{Adressa} & $\surd$        & $\surd$ & $\surd$   & $\surd$   &     &    &    & $\surd$        \\ \hline
\textbf{ZhihuRec}                 & $\surd$ & $\surd$ & $\surd$   & $\surd$   & $\surd$    & $\surd$ & $\surd$   & $\circ$    \\ \hline
\end{tabular}\label{tab:typeDifference}
\end{table*}

\section{Related Work}
Recommendation system \cite{lu2015recommender,adomavicius2005toward} has been a research hotspot in recent years. Some typical recommendation datasets have been widely used and contributed a lot to the research of recommender systems. One type of recommendation dataset mainly records the user's static information, such as ratings and favorites. The other type of dataset mainly records the user's dynamic behavior information, such as clicks and impressions.

Typical static recommendation datasets like movielens dataset\footnote{https://grouplens.org/datasets/movielens/} usually contain user-item interactions, timestamps and some metadata~\cite{2015movielens}. Old movielens dataset such as movielens-100k has small data size and contains only a few user proflies and movie attributes, recently movielens dataset such as movielens-20M has much larger data size and more movie attributes such as tags. Amazon dataset\footnote{http://jmcauley.ucsd.edu/data/amazon/} and Yelp\footnote{https://www.yelp.com/dataset/} dataset also contain textual content of user reviews~\cite{2016Julian,2010yelp,2015McAuley} . Another new dataset named Adressa\footnote{http://reclab.idi.ntnu.no/dataset/} ~\cite{2017Adressa} (in news recommendation scenario) provides useful information for the recommendation, while the dataset is a bit small (only 15514 users and 923 items are included).  These datasets usually contain information about ratings or favorites. Although there is no individual favorite relationship in the ZhihuRec dataset, the total number of personal favorites is recorded.

Recently, Xing dataset, released in RecSys Challenge 2017 \cite{abel2017recsys}, contains large-scale user click and impression logs in job recommendation scenarios and detailed content information. This dataset contains scarce impression information, which can be regarded as implicit negative feedback of users. The difference between us is that ZhihuRec provides rich text information from the questions and answers, and it also offers search queries.

Besides the classical news, e-commerce, and movie recommendation scenarios, knowledge sharing communities (Question \& Answer platform) have been widely used these years, such as Quora\footnote{https://www.quora.com} and Zhihu\footnote{https://www.zhihu.com}. Previous studies focus on natural language processing technology, such as reading comprehension and answer ranking, based on datasets (e.g.: SQuAD \cite{rajpurkar2016squad}, Question Answering Corpus \cite{nips15_hermann} and AQUA-RAT \cite{ling2017program}) in this scenario. With the rapid growth of recommendation and large-scale Q\&A pairs, Q\&A pair recommendation is necessary for users to provide them better services. Thus, we want to share a recommendation dataset with all user-item interactions and the content of users and items (Q\&A here).


To show the differences between ZhihuRec and previous recommendation studies, Table \ref{tab:typeDifference} is proposed. It shows that the ZhihuRec dataset contains more information and types than traditional recommendation datasets, including all information other datasets contain and some user query keywords.

\section{Dataset Construction}
\begin{figure}
\centering
\includegraphics[width=8.5cm]{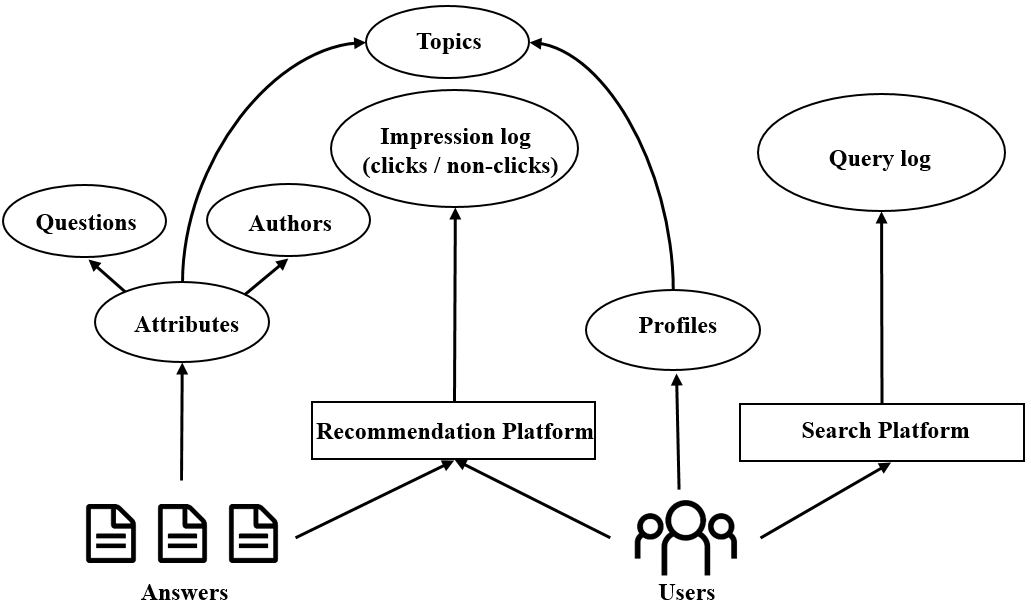}

\caption{ Dataset construction illustration}

\label{fig:dataconstruction}
\end{figure}
\subsection{Data Sampling}
The raw dataset is collected from May 3, 2018, to May 13, 2018. In the mobile scenario, over 1.3M users are involved. All the title and content information of users \& items (answers) are contained, and this dataset's language is Chinese. The users whose click number is less than 10 times are removed, and each user keeps his or her at most 160 latest interaction logs (including clicks and impressions). The average interaction number of each user is calculated, and we randomly sampled about 798k users to ensure the total user-answer interaction number is about 100M. User clicks are viewed as positive interaction, and user impressions (without clicks) are viewed as negative interaction. All user profiles are collected, and the attributes of all the topics of users are also collected. All answers the users interacted with are also collected, and from their attributes, all involved question attributes, author profiles, and topic attributes are also collected in the dataset. The process of ZhihuRec dataset construction is depicted in figure \ref{fig:dataconstruction}.

Besides, each user's at most 20 query logs are also recorded in the dataset. It is totally different from interaction logs. User's query keywords and the timestamp are recorded in the query log. The query keywords are very critical in information retrieval, and it is very convenient for cross-platform applications.

Table \ref{tab:impressionfieldDiscription} shows the fields of each impression record in ZhihuRec and their descriptions. All users' clicked and non-clicked impressions are recorded in the dataset according to answer read time. Table \ref{tab:queryfieldDiscription} shows the fields of each query record in the ZhihuRec dataset and their descriptions. All users' query keywords and the timestamp are recorded in the dataset.


\begin{table}[]
\centering
\caption{The impression fields of ZhihuRec dataset.}
\begin{tabular}{|l|p{5cm}|}
\hline
\textbf{Field}              & \textbf{Description}                                                   \\ \hline
userID             & Id of user                                                    \\ \hline
\#Impressions       & \#user-answer impression \\ \hline
$answerID_{i} $      & Id of the \textit{i}th answer \\ \hline
$show time_{i} $      & show time of the \textit{i}th answer \\ \hline
$read time_{i} $      & read time of the \textit{i}th answer \\ \hline

\end{tabular}\label{tab:impressionfieldDiscription}
\end{table}

\begin{table}[]
\centering
\caption{The query fields of the ZhihuRec dataset.}
\begin{tabular}{|l|p{5cm}|}
\hline
\textbf{Field}              & \textbf{Description}                                                   \\ \hline
userID             & Id of user                                                    \\ \hline
\#queries            & the total number of queries the user submitted           \\ \hline
$keywords_{i} $           & the keywords in the user's \textit{i}th query           \\ \hline
$query time_{i}  $          & query  timestamp of the user's \textit{i}th query         \\ \hline

\end{tabular}\label{tab:queryfieldDiscription}
\end{table}

As the ZhihuRec dataset contains about 100M user-answer interactions, it is also called Zhihu100M. Two smaller datasets randomly sampled from Zhihu100M dataset called Zhihu20M and Zhihu1M are also constructed to facilitate various application requirements. They contain about 20M and 1M user-answer impression logs and can be viewed as a medium-size dataset and a relatively small-size dataset. Some statistic information of them are displayed in Table~\ref{tab:statistic}.

\begin{table}[!h]
\caption{ZhihuRec Statistics}
\begin{tabular}{|l|c|c|c|}
\hline
\multicolumn{1}{|c|}{\textbf{Dataset}} & \textbf{Zhihu100M} & \multicolumn{1}{l|}{\textbf{Zhihu20M}} & \multicolumn{1}{l|}{\textbf{Zhihu1M}} \\ \hline
\#users & 798,086 & 159,878 & 7,963 \\ \hline
\#answers & 554,976 & 342,737 & 81,214 \\ \hline
\#impressions & 99,978,523 & 19,999,502 & 1,000,026 \\ \hline
\#queries & 3,900,243 & 779,104 & 38,148 \\ \hline
\#users with queries & 501,918 & 100,611 & 4,975 \\ \hline
\#clicks & 26,981,583 & 5,395,962 & 271,725 \\ \hline
avg  \#impressions & 125.27 & 125.09 & 125.58 \\ \hline
avg \#clicks & 33.81 & 33.75 & 34.12 \\ \hline
\#clicks : \#non-clicks& 1 : 2.71 & 1 : 2.71 & 1 : 2.68 \\ \hline
avg \#queries per user& 4.89 & 4.87 & 4.79 \\ \hline
\end{tabular}\label{tab:statistic}
\end{table}

\subsection{Data Characteristics}

User profiles and item attributes are all recorded in the ZhihuRec. This dataset retains the content information of users, questions, answer, and authors.
Table \ref{tab:userDiscription} shows the attributes of users, table \ref{tab:answerDiscription} shows the attributes of answers, table \ref{tab:questionDiscription} shows the attributes of questions and table \ref{tab:authorDiscription} shows the attributes of authors. As shown in Tables, there are abundant features about users, questions, answers, and authors comprehensively modeling both users and items (answers). There is no authorID in question attributes; the reason is many people can modify the questions in the Zhihu QA community over time. Notice that authorIDs are different from userIDs, which means if one person plays both roles of user and author in the dataset, his authorID and userID are not the same, as publisher and reader should not be shared.

\begin{table}[]
\centering
\caption{Main user profiles in ZhihuRec dataset, user ID is hashed and all text has been tranferred to ID, all information has been annoymized, including \textit{gender, login frequency, register type, province and city} }
\begin{tabular}{|c||c|}
\hline
\textbf{Demographics} & \textbf{QA Community related} \\ \hline
userID & \#followers \\ \hline
register time & \#follow topics \\ \hline
gender & \#follow questions \\ \hline
login frequency & \#answers \\ \hline
register type & \#questions \\ \hline
province & \#comments \\ \hline
city & \#get\_likes \\ \hline
 & \#get\_comments \\ \hline
 & \#get\_dislikes \\ \hline
 & followed topics \\ \hline
\end{tabular}\label{tab:userDiscription}
\end{table}

\begin{table}[]
\centering
\caption{Main answer attributes in ZhihuRec dataset, answer ID is hashed and all text has been tranferred to ID, all information has been annoymized}
\begin{tabular}{|c||c|}
\hline
\textbf{Basic Attributes} & \textbf{QA Community related} \\ \hline
answerID & \#get\_thanks \\ \hline
questionID & \#get\_likes \\ \hline
isAnonymous & \#get\_comments \\ \hline
authorID & \#get\_collections \\ \hline
isExcellentAnswer & \#get\_dislikes \\ \hline
isEditorRec & \# get\_reports \\ \hline
createTime &  \\ \hline
containPicture &  \\ \hline
containVideo &  \\ \hline
context &  \\ \hline
\end{tabular}\label{tab:answerDiscription}
\end{table}

\begin{table}[]
\centering
\caption{Main question attributes in ZhihuRec dataset, question ID is hashed and all text has been tranferred to ID, all information has been annoymized}
\begin{tabular}{|c||c|}
\hline
\textbf{Basic Attributes} & \textbf{QA Community related} \\ \hline
questionID & \#followers \\ \hline
createTime & \#answers \\ \hline
followed topics & \#invitations \\ \hline
title & \#comments \\ \hline
\end{tabular}\label{tab:questionDiscription}
\end{table}

\begin{table}[]
\centering
\caption{Main author profiles in ZhihuRec dataset, author ID is hashed, all information has been annoymized}
\begin{tabular}{|c||c|}
\hline
\textbf{Demographics}  & \textbf{QA Community related} \\ \hline
authorID   & \#followers\\ \hline
&isExcellentAuthor  \\ \hline
&isExcellentAnswerer  \\ \hline
\end{tabular}\label{tab:authorDiscription}
\end{table}

Each user or question also has several topic words (from 0 to 70,308), which are labeled by the user himself/herself (user topic words) or system users (question topic words, all users can edit them). It provides a more explicit way to help us understand user interests and the type of question, which is also useful for the recommendation. Each topic has a topic ID and topic description as its attributes, topic ID is hashed, and all context in topic description has been transferred to digital numbers. 

To the best of our knowledge, the ZhihuRec dataset is the largest public recommendation dataset collected from Zhihu online knowledge-sharing community containing almost all kinds of information such as abundant content information, user-answer interactions (clicks and impressions), user query logs, user profiles, answer attributes, question attributes, author profiles and topic attributes.


\subsection{Anonymization and Privacy Protection}
As the whole dataset is collected from a real scenario from real users, it is necessary to protect user privacy. Thus, not all content information of users is released.

All IDs in the ZhihuRec dataset are all anonymized and hashed. All text information such as the titles of questions, the content of the answers, the description of topics, and query keywords are splitted into words, and all the words are replaced by digital numbers. All the text features in user profiles (such as \textit{gender, register type, login frequency, province, city}) are all transferred to digital numbers, too. Thus the detailed information of user-profiles and article attributes can not be referenced from the ZhihuRec dataset.

Sensitive user information such as date of birth, work history, and education history has been removed. The network information of users (e.g., IP addresses) has also been removed.

Users' explicit feedback to answer such as \textit{like, thank, collection, comment, dislike and report} is hided, only statistic description to answers such as \#get\_like, \#get\_thank, \#get\_collection, \#get\_comment, \#get\_dislike and \#get\_report is provided.

\begin{figure}
\centering
\includegraphics[width=8cm]{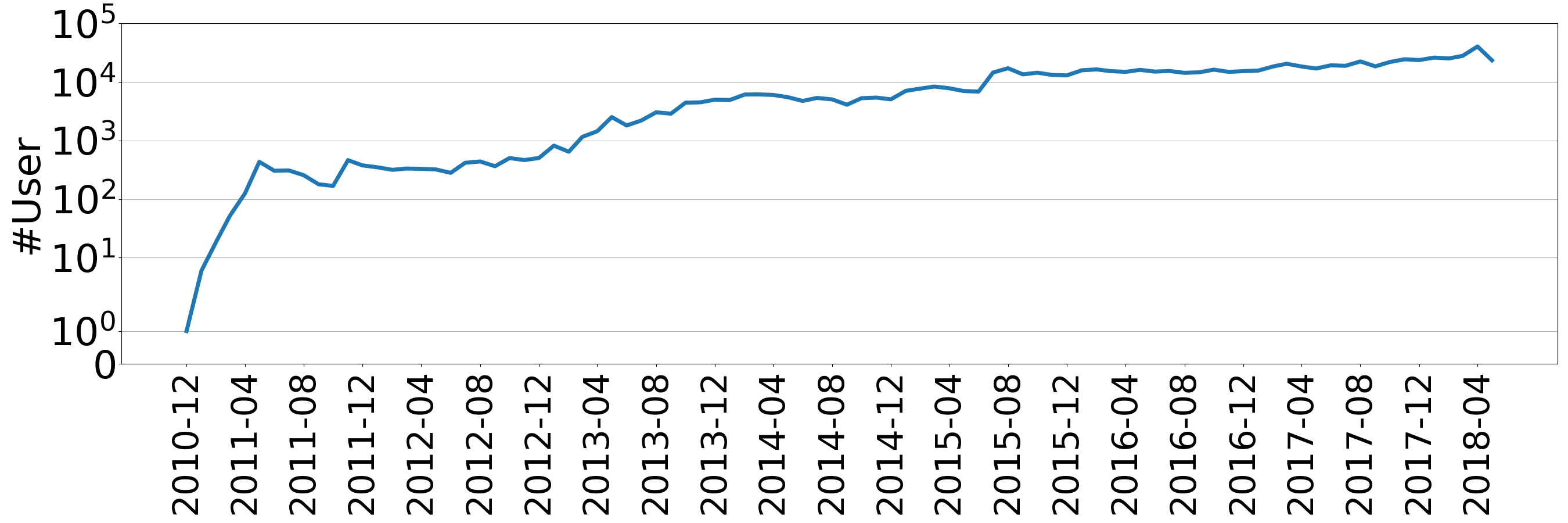}

\caption{ Distribution of user registration time}

\label{fig:useregistration}
\end{figure}

\begin{figure}
\centering
\includegraphics[width=8cm]{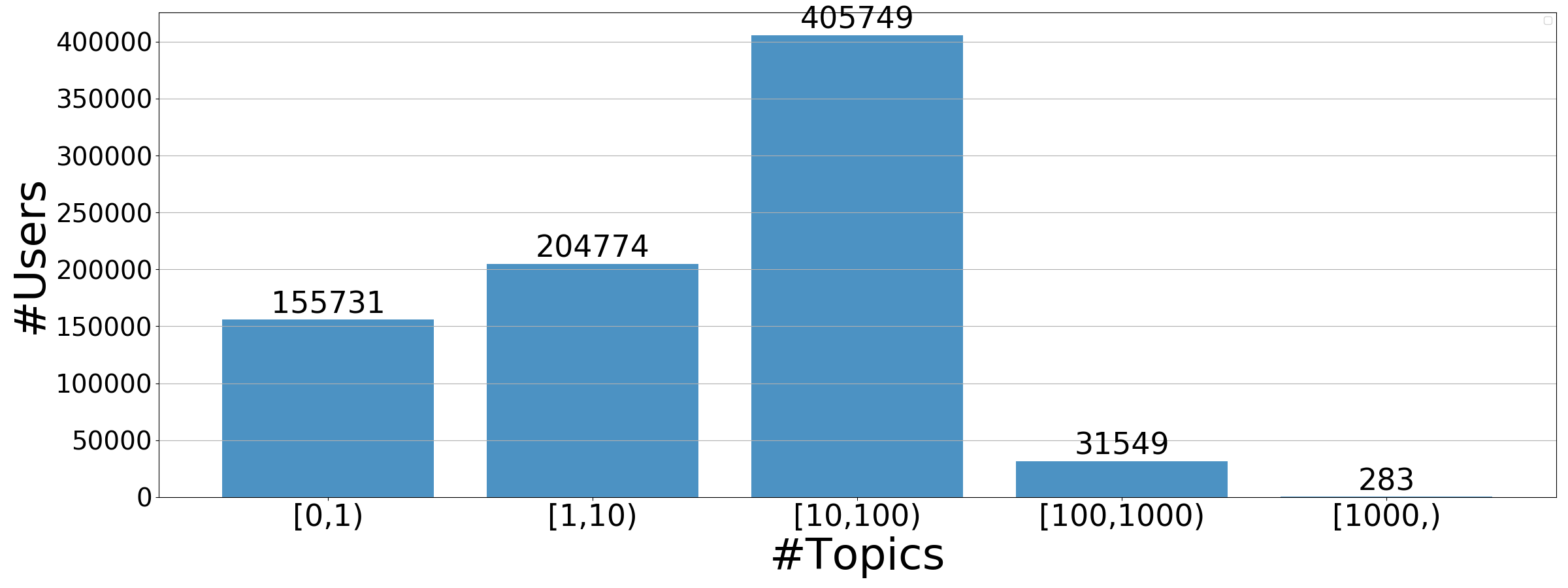}

\caption{ Distribution of users with different $\#$topics}

\label{fig:usertopic}
\end{figure}

\begin{figure}
\centering
\includegraphics[width=8cm]{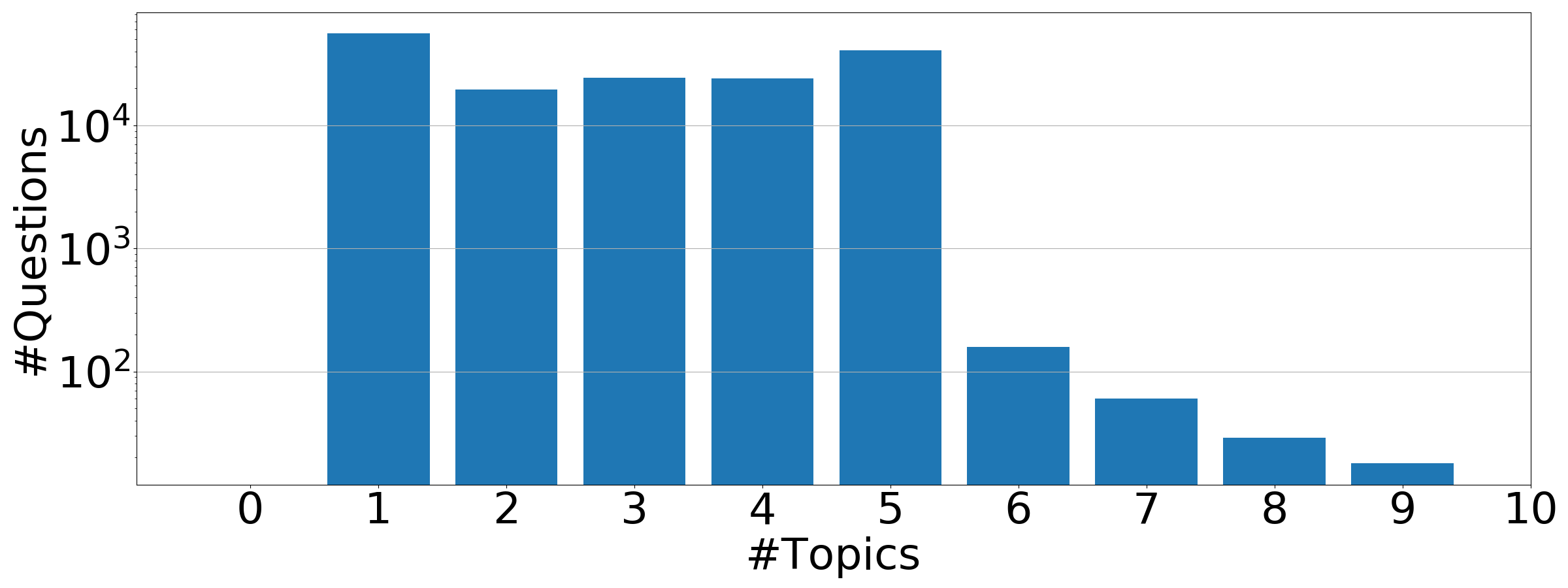}

\caption{ Distribution of questions with different $\#$topics, each question has at least one topic}

\label{fig:questiontopic}
\end{figure}

\begin{figure}
\centering
\includegraphics[width=8cm]{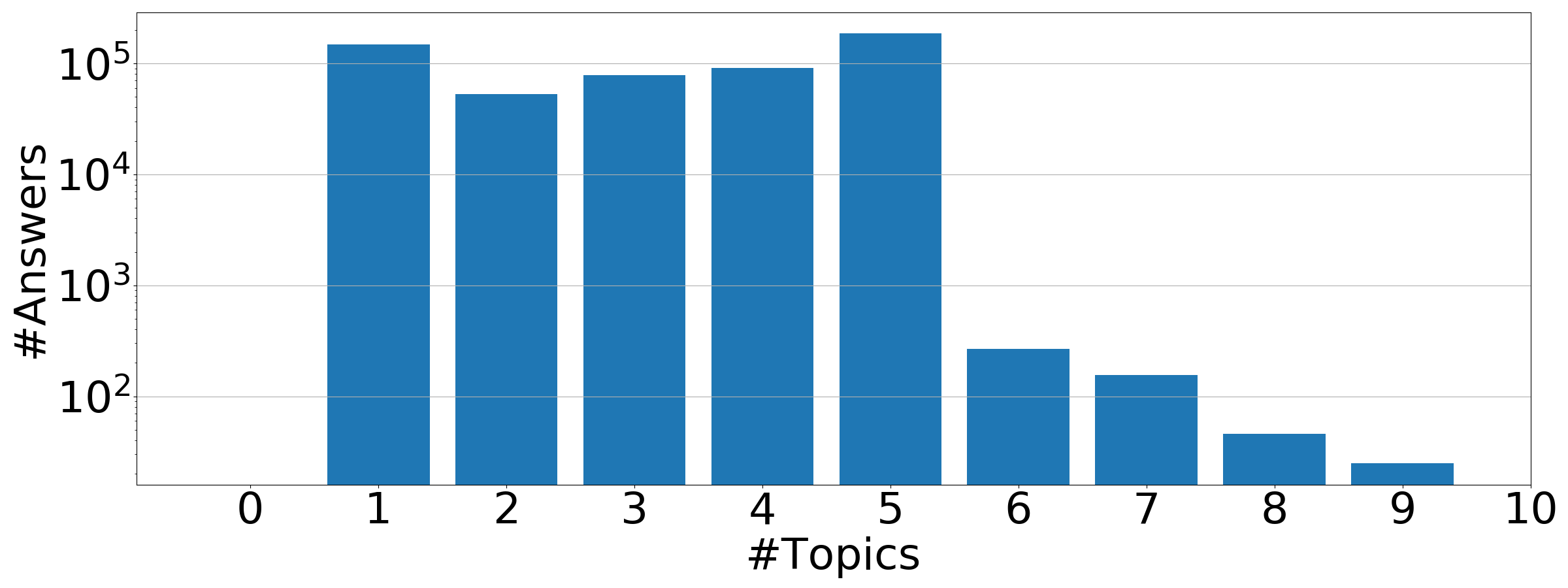}

\caption{ Distribution of answers with different $\#$topics, each answer has at least one topic}
\label{fig:answertopic}
\end{figure}

\begin{figure}
\centering
\includegraphics[width=8cm]{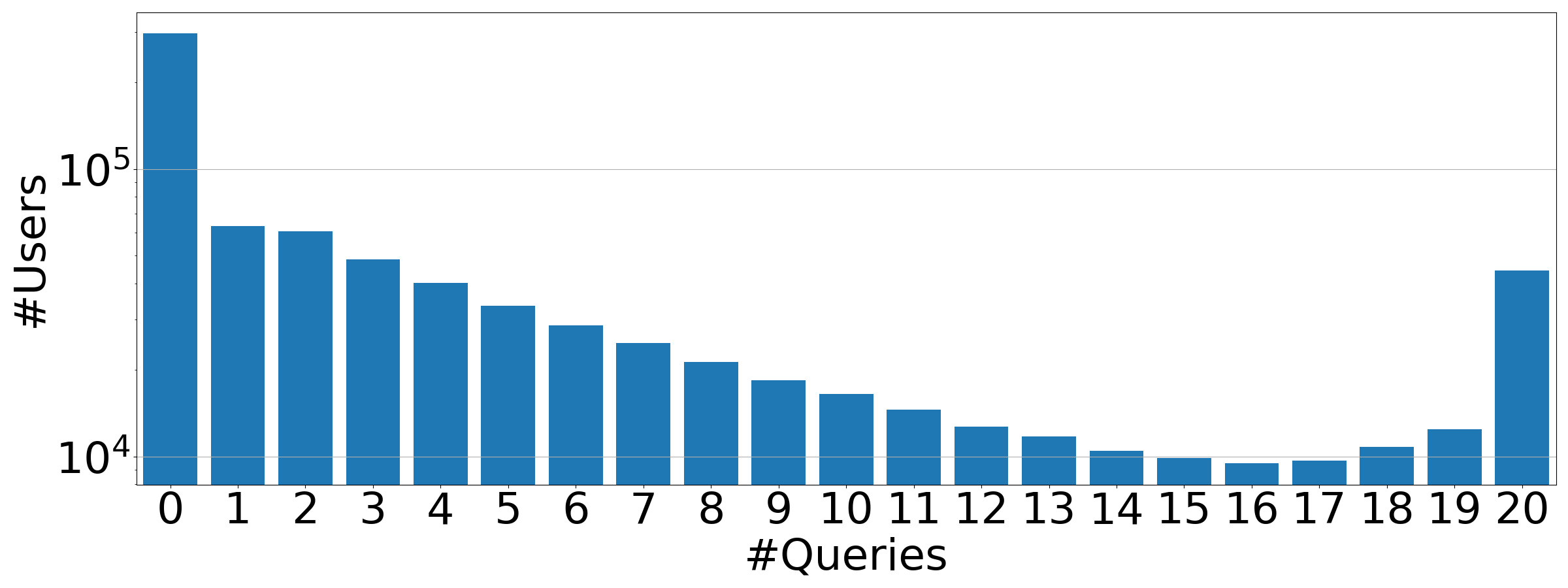}
\caption{ Distribution of users with different $\#$queries}
\label{fig:testsearch}
\end{figure}
\subsection{Statistic and Characteristic}
To better understand the ZhihuRec dataset scale, Table \ref{tab:statistic} provides some basic statistics. As outlined in the table, the dataset is much larger than the traditional recommendation dataset, closer to the training set for the real production environment.

Figure \ref{fig:useregistration} shows the distribution of user registration time; it can be found that the number of registered users per month gradually increases over time. Figure \ref{fig:usertopic} shows the number of user distribution per topic, figure \ref{fig:questiontopic} shows the number of question distribution per topic and figure \ref{fig:answertopic} shows the number of answer distribution per topic. It shows that most users follow on less than 100 topics, most answers, and most questions follow on more than one topic.

As search behavior is a new user interaction feature comparing with previous recommendation datasets,  Figure  \ref{fig:testsearch}  shows the number of  user distribution per query in the ZhihuRec dataset. Most users have less than 3 queries, and the distribution shows a log-like decay. However, there are many users with 20 queries, and the reason is that we did a truncation here (at most 20 recent query keywords of the user will be maintained).

\section{dataset application in multiple  Recommendation tasks }

\subsection{General Top N Recommendation}
Users' interaction logs are contained in the ZhihuRec dataset; from the perspective of the recommender system, the answers the users interacted in the logs can be viewed as items. This information is suitable for collaborative filtering \cite{thorat2015survey}, which contains the main methods in the general top N recommendation. To evaluate the performance of ZhihuRec dataset, 5 recommendation algorithms are applied in Zhihu1M dataset.

\begin{itemize}
    \item $\textbf{Pop}$ : This baseline always recommends the most popular answers (clicked by users) in the training set.
     \item $\textbf{ItemKNN}$\cite{itemknn2004}:  This method selects the top K nearest answer neighbors and uses their information to make predictions.
     \item $\textbf{BPR}$ \cite{rendleBPR2009}: This method applies the Bayesian Personalized Ranking objective function to optimize Matrix Factorization.
     \item $\textbf{LightGCN}$\cite{heLightGCN2020}: This method uses graph convolution network to enhance the performance of collaborative filtering.
     \item $\textbf{ENMF}$\cite{chenENMF2020}: An non-sampling neural recommendation model with efficient neural matrix factorization.
\end{itemize}

The experiment has been done with RecBole \cite{recbole}. The embedding size for user and answer is 64 for all methods. The neighbor number of ItemKNN is 100. Leave-one-out is adopted. Experimental results are shown in Table \ref{tab:genealRec}.

\begin{table}[!h]
\centering
\caption{Top-N Recommendation Results on Zhihu1M }
\begin{tabular}{|c|c|c|c|c|}
\hline
\textbf{Model} & \textbf{HR}@10 & \textbf{HR}@20 & \textbf{NDCG}@10 & \textbf{NDCG}@20\\ \hline
\textbf{Pop} & 0.0123  & 0.0198  & 0.0063  & 0.0081 \\ \hline
\textbf{ItemKNN} & 0.0129  & 0.0222  & 0.0063 & 0.0086 \\ \hline
\textbf{BPR} & 0.0225 & 0.0391 & 0.0110  & 0.0152 \\ \hline
\textbf{LightGCN} & 0.0259  & 0.0467  & 0.0122  & 0.0174 \\ \hline
\textbf{ENMF} & 0.0311  & 0.0543  & 0.0147  & 0.0210 \\ \hline

\end{tabular}\label{tab:genealRec}
\vspace{8pt}
\end{table}

\subsection{Sequential Recommendation}
Sequential recommendation play an important role in improving  many recommendation tasks' performance, as they reveal user's dynamic preferences. It is also a kind of top N recommendation. Usually, the difference between sequential recommendation and traditional recommendation is that sequential recommendation needs clear time information. It uses the item sequence interacted by the user as the input and sorts the item according to the interaction timestamp. It is also trending in the recommendation system. As all the users' interactions are logged in ZhihuRec dataset, in this paper, four state of art sequential models (FPMC \cite{rendleFPMC2010},  GRU4Rec\cite{hidasi2015session}, NARM\cite{liNARM2017}, SASRec \cite{kang2018self})have been applied in Zhihu1M dataset.

\begin{itemize}
    \item $\textbf{FPMC}$ \cite{rendleFPMC2010}: This method is based on personalized transition graphs over underlying Markov chains and combines MF.
     \item $\textbf{GRU4Rec}$\cite{hidasi2015session}:  A session-based model that uses RNN to capture sequential dependencies and make predictions.
     \item $\textbf{NARM}$ \cite{liNARM2017}: This method uses a hybrid encoder with an attention mechanism to capture the user's intention.
     \item $\textbf{SASRec}$\cite{kang2018self}: A sequential model that adopts the self-attention layer to capture the dynamic user interaction sequences.
\end{itemize}
The experiment has been done with RecBole \cite{recbole}.  The embedding size for user and answer is 64 for all methods. Leave-one-out is used. Experimental results are shown in Table \ref{tab:seqRec}.

\begin{table}[!h]
\centering
\caption{Sequential Recommendation Results on Zhihu1M}
\begin{tabular}{|c|c|c|c|c|}
\hline
\textbf{Model} & \textbf{HR}@10 & \textbf{HR}@20 & \textbf{NDCG}@10 & \textbf{NDCG}@20\\ \hline
\textbf{FPMC} & 0.0189  & 0.0339  & 0.0091  & 0.0129 \\ \hline
\textbf{GRU4Rec} & 0.0285  & 0.0501  & 0.0136 & 0.0191 \\ \hline
\textbf{NARM} & 0.0291 & 0.0510 & 0.0136  & 0.0191 \\ \hline
\textbf{SASRec} & 0.0314  & 0.0539  & 0.0152  & 0.0208 \\ \hline
\end{tabular}\label{tab:seqRec}
\vspace{8pt}
\end{table}

\subsection{Context-aware Recommendation}
Context-aware recommendation models use features from users, items, and context to enhance model performance \cite{verbert2012context,burke2002hybrid}. Context-aware recommendation combines the advantage of different recommendation models such as collaborative filtering, content-based model to get better recommendations; this dataset is very suitable for context-aware recommendation. As usually described in the click prediction task, the interaction representation by that one user clicks one answer is labeled to 1, and the interaction representation by that one user has an impression, but without clicking one answer is labeled to 0. As depicted in \cite{krichene2020sampled}, AUC (Area Under Curve, higher is better) is consistent under-sampling and is used as the evaluation metric. In this paper, 4 state-of-the-art context-aware models are applied in Zhihu1M dataset.

\begin{itemize}
    \item $\textbf{Wide\&Deep}$ \cite{cheng2016wide}:  Proposed by Google, which combines deep neural network and linear model, and widely used in real scenarios.
     \item $\textbf{NFM}$ \cite{he2017nfm}: A neural model that uses Bi-Interaction Layer to model second-order feature interactions.
     \item $\textbf{ACCM}$\cite{shi2018attention}: This is an attentional collaborative \& content model that unifies both content and user interactions.
     \item $\textbf{CC-CC}$\cite{shi2019adaptive}: This method uses the adaptive "Feature Sampling'' strategy for the recommendation.
\end{itemize}

The experiment has been done with CC-CC toolbox \footnote{https://github.com/THUIR/CC-CC}.  The embedding size of user and answer is 64 for all methods. For each user, the last click and  impressions after the last click are treated as the test set, the click right before the last click, and the impressions that happened between the click right before the last click and the last click are treated as the validation set, and others are treated as the training set. Experimental results are shown in Table \ref{tab:contextRec}.


\begin{table}[!h]
\centering
\caption{Context-aware Recommendation on Zhihu1M}
\begin{tabular}{|c|c|c|c|c|c|c|}
\hline
\multicolumn{1}{|c|}{\textbf{Model}} & \textbf{Wide\&Deep} & \textbf{NFM} & \textbf{ACCM} & \textbf{CC-CC}  \\ \hline
\textbf{AUC} & 0.6845 & 0.6843 & 0.6858 & 0.6901  \\ \hline
\end{tabular}\label{tab:contextRec}
\vspace{8pt}
\end{table}

\subsection{Discussion on Other Possible Recommendation Applications}
\subsubsection{Cross scenario Recommendation}
As mentioned above, users' query keywords are also included in the ZhihuRec dataset; the queries that the users have been searched for can be regarded as their explicit information need. While previous recommendation studies mainly focus on learning from users' implicit feedback. It will be helpful if more researchers try to integrate search and recommendation, which will help understand users' information needs better and provide better information services. The dataset can be applied to this kind of study due to its abundant query and recommendation logs.

\subsubsection{Recommendation with negative feedbacks}
When the users interact with answers, they give positive and negative feedbacks to them. Positive feedbacks are like the users' click, bookmark, and like the answers. Negative feedbacks are like the users delete and skip the answers. Traditional recommendation dataset lacks negative feedback, suffers from one-class problem \cite{pan2008one}. ZhihuRec dataset records both positive feedback and negative feedback of the users. As mentioned in \cite{paudel2018loss}, utilizing negative user preference can improve recommendation quality; this dataset is suitable for recommendation models with negative feedbacks.

\section{Tasks Beyond Recommendation}
Besides the usage for recommendation, we present two experiments of potential usage directions of the ZhihuRec dataset:

\subsection{Most Valuable Answerer Indentification}
We collected all the user profiles in the Zhihu100M dataset and named those users whose likes of his average answer are larger or equal to 50 as most valuable answerer and the others as non-most valuable answerer. The number of most valuable answerer is 22,818 (2.86\%), and the number of non-most valuable answerer is 775,268 (97.14\%).  The features are selected from the user to do this task are gender, login frequency, the number of users, questions and topics the user followed,  the number of the user's answers, questions and comments, the number of comments, supports, and against the user gets, register type, register platform. We use six models as our baselines: LinearSVC, Decision Tree (DT), Naive Bayes (NB), K-nearest neighbors (KNN), Random Forest (RF), Multilayer Perceptron (MLP). The toolbox we choose is sklearn. The evaluation metrics we used are accuracy, precision, recall, and f1 and 5-fold cross-validation in adopted. Results are shown in Table \ref{tab:mostvaluableanswererPrediction}.

\begin{table}[!h]
\centering
\caption{Most valuable answerer prediction on Zhihu1M.}
\begin{tabular}{|c|c|c|c|c|}
\hline
\textbf{Model} & \textbf{Accuracy} & \textbf{Precision} & \textbf{Recall} & \textbf{F1} \\ \hline
\textbf{KNN} & 0.9729 & 0.7003 & 0.0886 & 0.1573 \\ \hline
\textbf{Naive Bayes} & 0.8463 & 0.1097 & 0.6149 & 0.1861 \\ \hline
\textbf{LinearSVC} & 0.9783 & 0.7009 & 0.4179 & 0.5236 \\ \hline
\textbf{Decision Tree} & 0.9720 & 0.5091 & 0.5587 & 0.5327 \\ \hline
\textbf{Random Forest} & 0.9805 & 0.7981 & 0.4265 & 0.5559 \\ \hline
\textbf{MLP} & 0.9831 & 0.7666 & 0.5869 & 0.6648 \\ \hline

\end{tabular}\label{tab:mostvaluableanswererPrediction}
\vspace{8pt}
\end{table}

\subsection{High Quality Answer Recognition}
We collected all the answer features in the Zhihu100M dataset and named those answers which are recommended by the editor as high-quality answers and the others as non-high quality answers. The number of high-quality answers is 13,333 (2.40\%) and the number of non-high-quality answers is 541,643 (97.60\%). The features are selected from the answer to do this task are whether the answer is anonymous and excellent answer, whether the answer contains pictures and videos, the number of thanks, supports, comments, favorites, against and helpless the answer gets. We use six models as our baselines: SVM with the linear kernel (LinearSVC), Decision Tree (DT), Naive Bayes (NB), K-nearest neighbors (KNN), Random Forest (RF), Multilayer Perceptron with one hidden layer of 100 nodes (MLP). The toolbox we choose is sklearn. The evaluation metric we use is accuracy, precision, recall, and f1. We use 5-fold cross-validation and experimental results are shown in Table \ref{tab:highqualityanswerPrediction}.




\begin{table}[!h]
\centering
\caption{High quality answer prediction on Zhihu1M.}
\begin{tabular}{|c|c|c|c|c|}
\hline
\textbf{Model} & \textbf{Accuracy} & \textbf{Precision} & \textbf{Recall} & \textbf{F1} \\ \hline
\textbf{KNN} & 0.9753 & 0.4507 & 0.1270 & 0.1981 \\ \hline
\textbf{Naive Bayes} & 0.8871 & 0.1285 & 0.6402 & 0.2141 \\ \hline
\textbf{Decision Tree} & 0.9648 & 0.3070 & 0.3699 & 0.3355 \\ \hline
\textbf{LinearSVC} & 0.9809 & 0.9475 & 0.2152 & 0.3507 \\ \hline
\textbf{Random Forest} & 0.9804 & 0.7744 & 0.2614 & 0.3908 \\ \hline
\textbf{MLP} & 0.9810 & 0.8474 & 0.2544 & 0.3913 \\ \hline
\end{tabular}\label{tab:highqualityanswerPrediction}
\vspace{8pt}
\end{table}

\section{CONCLUSIONS}
This paper presents a new dataset of the online knowledge-sharing community, which aims to contribute to personalized recommendations. To the best of our knowledge, it is the largest dataset with detailed content features, including users, items, impressions, authors, topics, and user interaction logs with search queries and clicks on recommendation results. Experimental results on state-of-art algorithms have been depicted. This dataset is feasible to be used in researches on context-aware recommendation, sequential recommendation, recommendation with negative feedback, integrating search and recommendation, and user profiling and item attribute modeling. This dataset is publicly available and contains abundant information in interaction logs and query keywords which is suitable for cross-platform researches.
\clearpage

\bibliographystyle{ACM-Reference-Format}
\bibliography{zhihu_resource}


\begin{thebibliography}{31}


\ifx \showCODEN    \undefined \def \showCODEN     #1{\unskip}     \fi
\ifx \showDOI      \undefined \def \showDOI       #1{#1}\fi
\ifx \showISBNx    \undefined \def \showISBNx     #1{\unskip}     \fi
\ifx \showISBNxiii \undefined \def \showISBNxiii  #1{\unskip}     \fi
\ifx \showISSN     \undefined \def \showISSN      #1{\unskip}     \fi
\ifx \showLCCN     \undefined \def \showLCCN      #1{\unskip}     \fi
\ifx \shownote     \undefined \def \shownote      #1{#1}          \fi
\ifx \showarticletitle \undefined \def \showarticletitle #1{#1}   \fi
\ifx \showURL      \undefined \def \showURL       {\relax}        \fi
\providecommand\bibfield[2]{#2}
\providecommand\bibinfo[2]{#2}
\providecommand\natexlab[1]{#1}
\providecommand\showeprint[2][]{arXiv:#2}

\bibitem[\protect\citeauthoryear{Abel, Deldjoo, Elahi, and Kohlsdorf}{Abel
  et~al\mbox{.}}{2017}]%
        {abel2017recsys}
\bibfield{author}{\bibinfo{person}{Fabian Abel}, \bibinfo{person}{Yashar
  Deldjoo}, \bibinfo{person}{Mehdi Elahi}, {and} \bibinfo{person}{Daniel
  Kohlsdorf}.} \bibinfo{year}{2017}\natexlab{}.
\newblock \showarticletitle{Recsys challenge 2017: Offline and online
  evaluation}. In \bibinfo{booktitle}{\emph{ACM RecSys2017}}. ACM,
  \bibinfo{pages}{372--373}.
\newblock


\bibitem[\protect\citeauthoryear{Adomavicius and Tuzhilin}{Adomavicius and
  Tuzhilin}{2005}]%
        {adomavicius2005toward}
\bibfield{author}{\bibinfo{person}{Gediminas Adomavicius} {and}
  \bibinfo{person}{Alexander Tuzhilin}.} \bibinfo{year}{2005}\natexlab{}.
\newblock \showarticletitle{Toward the next generation of recommender systems:
  A survey of the state-of-the-art and possible extensions}.
\newblock \bibinfo{journal}{\emph{IEEE TKDE}} \bibinfo{number}{6}
  (\bibinfo{year}{2005}), \bibinfo{pages}{734--749}.
\newblock


\bibitem[\protect\citeauthoryear{Burke}{Burke}{2002}]%
        {burke2002hybrid}
\bibfield{author}{\bibinfo{person}{Robin Burke}.}
  \bibinfo{year}{2002}\natexlab{}.
\newblock \showarticletitle{Hybrid recommender systems: Survey and
  experiments}.
\newblock \bibinfo{journal}{\emph{UMUAI}} \bibinfo{volume}{12},
  \bibinfo{number}{4} (\bibinfo{year}{2002}), \bibinfo{pages}{331--370}.
\newblock


\bibitem[\protect\citeauthoryear{Chen, Zhang, Zhang, Liu, and Ma}{Chen
  et~al\mbox{.}}{2020}]%
        {chenENMF2020}
\bibfield{author}{\bibinfo{person}{Chong Chen}, \bibinfo{person}{Min Zhang},
  \bibinfo{person}{Yongfeng Zhang}, \bibinfo{person}{Yiqun Liu}, {and}
  \bibinfo{person}{Shaoping Ma}.} \bibinfo{year}{2020}\natexlab{}.
\newblock \showarticletitle{Efficient Neural Matrix Factorization without
  Sampling for Recommendation}.
\newblock \bibinfo{journal}{\emph{ACM Trans. Inf. Syst.}} \bibinfo{volume}{38},
  \bibinfo{number}{2}, Article \bibinfo{articleno}{14} (\bibinfo{date}{Jan.}
  \bibinfo{year}{2020}), \bibinfo{numpages}{28}~pages.
\newblock
\showISSN{1046-8188}
\urldef\tempurl%
\url{https://doi.org/10.1145/3373807}
\showDOI{\tempurl}


\bibitem[\protect\citeauthoryear{Cheng, Koc, and Harmsen}{Cheng
  et~al\mbox{.}}{2016}]%
        {cheng2016wide}
\bibfield{author}{\bibinfo{person}{Heng-Tze Cheng}, \bibinfo{person}{Levent
  Koc}, {and} \bibinfo{person}{Jeremiah et~al. Harmsen}.}
  \bibinfo{year}{2016}\natexlab{}.
\newblock \showarticletitle{Wide \& deep learning for recommender systems}. In
  \bibinfo{booktitle}{\emph{Proceedings of the 1st Workshop on Deep Learning
  for Recommender Systems}}. ACM, \bibinfo{pages}{7--10}.
\newblock


\bibitem[\protect\citeauthoryear{Deshpande and Karypis}{Deshpande and
  Karypis}{2004}]%
        {itemknn2004}
\bibfield{author}{\bibinfo{person}{Mukund Deshpande} {and}
  \bibinfo{person}{George Karypis}.} \bibinfo{year}{2004}\natexlab{}.
\newblock \showarticletitle{Item-Based Top-N Recommendation Algorithms}.
\newblock \bibinfo{journal}{\emph{ACM Trans. Inf. Syst.}} \bibinfo{volume}{22},
  \bibinfo{number}{1} (\bibinfo{date}{Jan.} \bibinfo{year}{2004}),
  \bibinfo{pages}{143–177}.
\newblock
\showISSN{1046-8188}
\urldef\tempurl%
\url{https://doi.org/10.1145/963770.963776}
\showDOI{\tempurl}


\bibitem[\protect\citeauthoryear{Gulla, Zhang, Liu, \"{O}zg\"{o}bek, and
  Su}{Gulla et~al\mbox{.}}{2017}]%
        {2017Adressa}
\bibfield{author}{\bibinfo{person}{Jon~Atle Gulla}, \bibinfo{person}{Lemei
  Zhang}, \bibinfo{person}{Peng Liu}, \bibinfo{person}{\"{O}zlem
  \"{O}zg\"{o}bek}, {and} \bibinfo{person}{Xiaomeng Su}.}
  \bibinfo{year}{2017}\natexlab{}.
\newblock \showarticletitle{The Adressa Dataset for News Recommendation}. In
  \bibinfo{booktitle}{\emph{Proceedings of the International Conference on Web
  Intelligence}} (Leipzig, Germany) \emph{(\bibinfo{series}{WI ’17})}.
  \bibinfo{publisher}{Association for Computing Machinery},
  \bibinfo{address}{New York, NY, USA}, \bibinfo{pages}{1042–1048}.
\newblock
\showISBNx{9781450349512}


\bibitem[\protect\citeauthoryear{Harper and Konstan}{Harper and
  Konstan}{2015}]%
        {2015movielens}
\bibfield{author}{\bibinfo{person}{Franklin Harper} {and}
  \bibinfo{person}{Joseph Konstan}.} \bibinfo{year}{2015}\natexlab{}.
\newblock \showarticletitle{The MovieLens Datasets}.
\newblock \bibinfo{journal}{\emph{ACM Transactions on Interactive Intelligent
  Systems}}  \bibinfo{volume}{5} (\bibinfo{date}{12} \bibinfo{year}{2015}),
  \bibinfo{pages}{1--19}.
\newblock
\urldef\tempurl%
\url{https://doi.org/10.1145/2827872}
\showDOI{\tempurl}


\bibitem[\protect\citeauthoryear{He and McAuley}{He and McAuley}{2016}]%
        {2016Julian}
\bibfield{author}{\bibinfo{person}{Ruining He} {and} \bibinfo{person}{Julian
  McAuley}.} \bibinfo{year}{2016}\natexlab{}.
\newblock \showarticletitle{Ups and Downs: Modeling the Visual Evolution of
  Fashion Trends with One-Class Collaborative Filtering}. In
  \bibinfo{booktitle}{\emph{Proceedings of the 25th International Conference on
  World Wide Web}} (Montr\'{e}al, Qu\'{e}bec, Canada)
  \emph{(\bibinfo{series}{WWW ’16})}. \bibinfo{publisher}{International World
  Wide Web Conferences Steering Committee}, \bibinfo{address}{Republic and
  Canton of Geneva, CHE}, \bibinfo{pages}{507–517}.
\newblock
\showISBNx{9781450341431}
\urldef\tempurl%
\url{https://doi.org/10.1145/2872427.2883037}
\showDOI{\tempurl}


\bibitem[\protect\citeauthoryear{He and Chua}{He and Chua}{2017}]%
        {he2017nfm}
\bibfield{author}{\bibinfo{person}{Xiangnan He} {and} \bibinfo{person}{Tat-Seng
  Chua}.} \bibinfo{year}{2017}\natexlab{}.
\newblock \showarticletitle{Neural factorization machines for sparse predictive
  analytics}. In \bibinfo{booktitle}{\emph{Proceedings of SIGIR2017}}. ACM,
  \bibinfo{pages}{355--364}.
\newblock


\bibitem[\protect\citeauthoryear{He, Deng, Wang, Li, Zhang, and Wang}{He
  et~al\mbox{.}}{2020}]%
        {heLightGCN2020}
\bibfield{author}{\bibinfo{person}{Xiangnan He}, \bibinfo{person}{Kuan Deng},
  \bibinfo{person}{Xiang Wang}, \bibinfo{person}{Yan Li},
  \bibinfo{person}{YongDong Zhang}, {and} \bibinfo{person}{Meng Wang}.}
  \bibinfo{year}{2020}\natexlab{}.
\newblock \bibinfo{booktitle}{\emph{LightGCN: Simplifying and Powering Graph
  Convolution Network for Recommendation}}.
\newblock \bibinfo{publisher}{Association for Computing Machinery},
  \bibinfo{address}{New York, NY, USA}, \bibinfo{pages}{639–648}.
\newblock
\showISBNx{9781450380164}
\urldef\tempurl%
\url{https://doi.org/10.1145/3397271.3401063}
\showURL{%
\tempurl}


\bibitem[\protect\citeauthoryear{Hermann, Ko\v{c}isk\'y, Grefenstette,
  Espeholt, Kay, Suleyman, and Blunsom}{Hermann et~al\mbox{.}}{2015}]%
        {nips15_hermann}
\bibfield{author}{\bibinfo{person}{Karl~Moritz Hermann},
  \bibinfo{person}{Tom\'a\v{s} Ko\v{c}isk\'y}, \bibinfo{person}{Edward
  Grefenstette}, \bibinfo{person}{Lasse Espeholt}, \bibinfo{person}{Will Kay},
  \bibinfo{person}{Mustafa Suleyman}, {and} \bibinfo{person}{Phil Blunsom}.}
  \bibinfo{year}{2015}\natexlab{}.
\newblock \showarticletitle{Teaching Machines to Read and Comprehend}. In
  \bibinfo{booktitle}{\emph{NIPS2015}}.
\newblock
\urldef\tempurl%
\url{http://arxiv.org/abs/1506.03340}
\showURL{%
\tempurl}


\bibitem[\protect\citeauthoryear{Hidasi, Karatzoglou, Baltrunas, and
  Tikk}{Hidasi et~al\mbox{.}}{2015}]%
        {hidasi2015session}
\bibfield{author}{\bibinfo{person}{Bal{\'a}zs Hidasi},
  \bibinfo{person}{Alexandros Karatzoglou}, \bibinfo{person}{Linas Baltrunas},
  {and} \bibinfo{person}{Domonkos Tikk}.} \bibinfo{year}{2015}\natexlab{}.
\newblock \showarticletitle{Session-based recommendations with recurrent neural
  networks}.
\newblock \bibinfo{journal}{\emph{arXiv preprint arXiv:1511.06939}}
  (\bibinfo{year}{2015}).
\newblock


\bibitem[\protect\citeauthoryear{Kang and McAuley}{Kang and McAuley}{2018}]%
        {kang2018self}
\bibfield{author}{\bibinfo{person}{Wang-Cheng Kang} {and}
  \bibinfo{person}{Julian McAuley}.} \bibinfo{year}{2018}\natexlab{}.
\newblock \showarticletitle{Self-attentive sequential recommendation}. In
  \bibinfo{booktitle}{\emph{2018 IEEE International Conference on Data Mining
  (ICDM)}}. IEEE, \bibinfo{pages}{197--206}.
\newblock


\bibitem[\protect\citeauthoryear{Krichene and Rendle}{Krichene and
  Rendle}{2020}]%
        {krichene2020sampled}
\bibfield{author}{\bibinfo{person}{Walid Krichene} {and}
  \bibinfo{person}{Steffen Rendle}.} \bibinfo{year}{2020}\natexlab{}.
\newblock \showarticletitle{On Sampled Metrics for Item Recommendation}. In
  \bibinfo{booktitle}{\emph{Proceedings of the 26th ACM SIGKDD International
  Conference on Knowledge Discovery \& Data Mining}}.
  \bibinfo{pages}{1748--1757}.
\newblock


\bibitem[\protect\citeauthoryear{Li, Ren, Chen, Ren, Lian, and Ma}{Li
  et~al\mbox{.}}{2017}]%
        {liNARM2017}
\bibfield{author}{\bibinfo{person}{Jing Li}, \bibinfo{person}{Pengjie Ren},
  \bibinfo{person}{Zhumin Chen}, \bibinfo{person}{Zhaochun Ren},
  \bibinfo{person}{Tao Lian}, {and} \bibinfo{person}{Jun Ma}.}
  \bibinfo{year}{2017}\natexlab{}.
\newblock \showarticletitle{Neural Attentive Session-Based Recommendation}. In
  \bibinfo{booktitle}{\emph{Proceedings of the 2017 ACM on Conference on
  Information and Knowledge Management}} (Singapore, Singapore)
  \emph{(\bibinfo{series}{CIKM '17})}. \bibinfo{publisher}{Association for
  Computing Machinery}, \bibinfo{address}{New York, NY, USA},
  \bibinfo{pages}{1419–1428}.
\newblock
\showISBNx{9781450349185}
\urldef\tempurl%
\url{https://doi.org/10.1145/3132847.3132926}
\showDOI{\tempurl}


\bibitem[\protect\citeauthoryear{Li, Nie, Zhang, Wang, Yan, and Weng}{Li
  et~al\mbox{.}}{2010}]%
        {2010yelp}
\bibfield{author}{\bibinfo{person}{Yize Li}, \bibinfo{person}{Jiazhong Nie},
  \bibinfo{person}{Yi Zhang}, \bibinfo{person}{Bingqing Wang},
  \bibinfo{person}{Baoshi Yan}, {and} \bibinfo{person}{Fuliang Weng}.}
  \bibinfo{year}{2010}\natexlab{}.
\newblock \showarticletitle{Contextual Recommendation based on Text Mining}. In
  \bibinfo{booktitle}{\emph{Coling 2010: Posters}}. \bibinfo{publisher}{Coling
  2010 Organizing Committee}, \bibinfo{address}{Beijing, China},
  \bibinfo{pages}{692--700}.
\newblock
\urldef\tempurl%
\url{https://www.aclweb.org/anthology/C10-2079}
\showURL{%
\tempurl}


\bibitem[\protect\citeauthoryear{Ling, Yogatama, Dyer, and Blunsom}{Ling
  et~al\mbox{.}}{2017}]%
        {ling2017program}
\bibfield{author}{\bibinfo{person}{Wang Ling}, \bibinfo{person}{Dani Yogatama},
  \bibinfo{person}{Chris Dyer}, {and} \bibinfo{person}{Phil Blunsom}.}
  \bibinfo{year}{2017}\natexlab{}.
\newblock \showarticletitle{Program induction by rationale generation: Learning
  to solve and explain algebraic word problems}.
\newblock \bibinfo{journal}{\emph{arXiv preprint arXiv:1705.04146}}
  (\bibinfo{year}{2017}).
\newblock


\bibitem[\protect\citeauthoryear{Lu, Wu, Mao, Wang, and Zhang}{Lu
  et~al\mbox{.}}{2015}]%
        {lu2015recommender}
\bibfield{author}{\bibinfo{person}{Jie Lu}, \bibinfo{person}{Dianshuang Wu},
  \bibinfo{person}{Mingsong Mao}, \bibinfo{person}{Wei Wang}, {and}
  \bibinfo{person}{Guangquan Zhang}.} \bibinfo{year}{2015}\natexlab{}.
\newblock \showarticletitle{Recommender system application developments: a
  survey}.
\newblock \bibinfo{journal}{\emph{Decision Support Systems}}
  \bibinfo{volume}{74} (\bibinfo{year}{2015}), \bibinfo{pages}{12--32}.
\newblock


\bibitem[\protect\citeauthoryear{McAuley, Targett, Shi, and van~den
  Hengel}{McAuley et~al\mbox{.}}{2015}]%
        {2015McAuley}
\bibfield{author}{\bibinfo{person}{Julian McAuley},
  \bibinfo{person}{Christopher Targett}, \bibinfo{person}{Qinfeng Shi}, {and}
  \bibinfo{person}{Anton van~den Hengel}.} \bibinfo{year}{2015}\natexlab{}.
\newblock \showarticletitle{Image-Based Recommendations on Styles and
  Substitutes} \emph{(\bibinfo{series}{SIGIR ’15})}.
  \bibinfo{publisher}{Association for Computing Machinery},
  \bibinfo{address}{SIGIR 2015}, \bibinfo{pages}{43–52}.
\newblock
\showISBNx{9781450336215}


\bibitem[\protect\citeauthoryear{Pan, Zhou, Cao, and Liu}{Pan
  et~al\mbox{.}}{2008}]%
        {pan2008one}
\bibfield{author}{\bibinfo{person}{Rong Pan}, \bibinfo{person}{Yunhong Zhou},
  \bibinfo{person}{Bin Cao}, {and} \bibinfo{person}{Nathan N et~al. Liu}.}
  \bibinfo{year}{2008}\natexlab{}.
\newblock \showarticletitle{One-class collaborative filtering}. In
  \bibinfo{booktitle}{\emph{ICDM 2008}}. IEEE, \bibinfo{pages}{502--511}.
\newblock


\bibitem[\protect\citeauthoryear{Paudel, Luck, and Bernstein}{Paudel
  et~al\mbox{.}}{2018}]%
        {paudel2018loss}
\bibfield{author}{\bibinfo{person}{Bibek Paudel}, \bibinfo{person}{Sandro
  Luck}, {and} \bibinfo{person}{Abraham Bernstein}.}
  \bibinfo{year}{2018}\natexlab{}.
\newblock \showarticletitle{Loss Aversion in Recommender Systems: Utilizing
  Negative User Preference to Improve Recommendation Quality}.
\newblock \bibinfo{journal}{\emph{arXiv preprint arXiv:1812.11422}}
  (\bibinfo{year}{2018}).
\newblock


\bibitem[\protect\citeauthoryear{Rajpurkar, Zhang, Lopyrev, and
  Liang}{Rajpurkar et~al\mbox{.}}{2016}]%
        {rajpurkar2016squad}
\bibfield{author}{\bibinfo{person}{Pranav Rajpurkar}, \bibinfo{person}{Jian
  Zhang}, \bibinfo{person}{Konstantin Lopyrev}, {and} \bibinfo{person}{Percy
  Liang}.} \bibinfo{year}{2016}\natexlab{}.
\newblock \showarticletitle{Squad: 100,000+ questions for machine comprehension
  of text}.
\newblock \bibinfo{journal}{\emph{arXiv preprint arXiv:1606.05250}}
  (\bibinfo{year}{2016}).
\newblock


\bibitem[\protect\citeauthoryear{Rendle, Freudenthaler, Gantner, and
  Schmidt-Thieme}{Rendle et~al\mbox{.}}{2009}]%
        {rendleBPR2009}
\bibfield{author}{\bibinfo{person}{Steffen Rendle}, \bibinfo{person}{Christoph
  Freudenthaler}, \bibinfo{person}{Zeno Gantner}, {and} \bibinfo{person}{Lars
  Schmidt-Thieme}.} \bibinfo{year}{2009}\natexlab{}.
\newblock \showarticletitle{BPR: Bayesian Personalized Ranking from Implicit
  Feedback}. In \bibinfo{booktitle}{\emph{Proceedings of the Twenty-Fifth
  Conference on Uncertainty in Artificial Intelligence}} (Montreal, Quebec,
  Canada) \emph{(\bibinfo{series}{UAI '09})}. \bibinfo{publisher}{AUAI Press},
  \bibinfo{address}{Arlington, Virginia, USA}, \bibinfo{pages}{452–461}.
\newblock
\showISBNx{9780974903958}


\bibitem[\protect\citeauthoryear{Rendle, Freudenthaler, and
  Schmidt-Thieme}{Rendle et~al\mbox{.}}{2010}]%
        {rendleFPMC2010}
\bibfield{author}{\bibinfo{person}{Steffen Rendle}, \bibinfo{person}{Christoph
  Freudenthaler}, {and} \bibinfo{person}{Lars Schmidt-Thieme}.}
  \bibinfo{year}{2010}\natexlab{}.
\newblock \showarticletitle{Factorizing Personalized Markov Chains for
  Next-Basket Recommendation}. In \bibinfo{booktitle}{\emph{Proceedings of the
  19th International Conference on World Wide Web}} (Raleigh, North Carolina,
  USA) \emph{(\bibinfo{series}{WWW '10})}. \bibinfo{publisher}{Association for
  Computing Machinery}, \bibinfo{address}{New York, NY, USA},
  \bibinfo{pages}{811–820}.
\newblock
\showISBNx{9781605587998}
\urldef\tempurl%
\url{https://doi.org/10.1145/1772690.1772773}
\showDOI{\tempurl}


\bibitem[\protect\citeauthoryear{Shaoyun~Shi and Ma}{Shaoyun~Shi and
  Ma}{2019}]%
        {shi2019adaptive}
\bibfield{author}{\bibinfo{person}{Xinxing Yu Yongfeng Zhang Bin Hao Yiqun~Liu
  Shaoyun~Shi, Min~Zhang} {and} \bibinfo{person}{Shaoping Ma}.}
  \bibinfo{year}{2019}\natexlab{}.
\newblock \showarticletitle{Adaptive Feature Sampling for Recommendation with
  Missing Content Feature Values}. In \bibinfo{booktitle}{\emph{Proceedings of
  the 28th ACM International Conference on Information and Knowledge
  Management}}. ACM, \bibinfo{pages}{1451--1460}.
\newblock


\bibitem[\protect\citeauthoryear{Sharma and Gera}{Sharma and Gera}{2013}]%
        {sharma2013survey}
\bibfield{author}{\bibinfo{person}{Lalita Sharma} {and} \bibinfo{person}{Anju
  Gera}.} \bibinfo{year}{2013}\natexlab{}.
\newblock \showarticletitle{A survey of recommendation system: Research
  challenges}.
\newblock \bibinfo{journal}{\emph{International Journal of Engineering Trends
  and Technology (IJETT)}} \bibinfo{volume}{4}, \bibinfo{number}{5}
  (\bibinfo{year}{2013}), \bibinfo{pages}{1989--1992}.
\newblock


\bibitem[\protect\citeauthoryear{Shi, Zhang, Liu, and Ma}{Shi
  et~al\mbox{.}}{2018}]%
        {shi2018attention}
\bibfield{author}{\bibinfo{person}{Shaoyun Shi}, \bibinfo{person}{Min Zhang},
  \bibinfo{person}{Yiqun Liu}, {and} \bibinfo{person}{Shaoping Ma}.}
  \bibinfo{year}{2018}\natexlab{}.
\newblock \showarticletitle{Attention-based Adaptive Model to Unify Warm and
  Cold Starts Recommendation}. In \bibinfo{booktitle}{\emph{Proceedings of
  CIKM2018}}. ACM, \bibinfo{pages}{127--136}.
\newblock


\bibitem[\protect\citeauthoryear{Thorat, Goudar, and Barve}{Thorat
  et~al\mbox{.}}{2015}]%
        {thorat2015survey}
\bibfield{author}{\bibinfo{person}{Poonam~B Thorat}, \bibinfo{person}{RM
  Goudar}, {and} \bibinfo{person}{Sunita Barve}.}
  \bibinfo{year}{2015}\natexlab{}.
\newblock \showarticletitle{Survey on collaborative filtering, content-based
  filtering and hybrid recommendation system}.
\newblock \bibinfo{journal}{\emph{International Journal of Computer
  Applications}} \bibinfo{volume}{110}, \bibinfo{number}{4}
  (\bibinfo{year}{2015}), \bibinfo{pages}{31--36}.
\newblock


\bibitem[\protect\citeauthoryear{Verbert, Manouselis, Ochoa, Wolpers,
  Drachsler, Bosnic, and Duval}{Verbert et~al\mbox{.}}{2012}]%
        {verbert2012context}
\bibfield{author}{\bibinfo{person}{Katrien Verbert}, \bibinfo{person}{Nikos
  Manouselis}, \bibinfo{person}{Xavier Ochoa}, \bibinfo{person}{Martin
  Wolpers}, \bibinfo{person}{Hendrik Drachsler}, \bibinfo{person}{Ivana
  Bosnic}, {and} \bibinfo{person}{Erik Duval}.}
  \bibinfo{year}{2012}\natexlab{}.
\newblock \showarticletitle{Context-aware recommender systems for learning: a
  survey and future challenges}.
\newblock \bibinfo{journal}{\emph{IEEE Transactions on Learning Technologies}}
  \bibinfo{volume}{5}, \bibinfo{number}{4} (\bibinfo{year}{2012}),
  \bibinfo{pages}{318--335}.
\newblock


\bibitem[\protect\citeauthoryear{Zhao, Mu, Hou, Lin, Li, Chen, Lu, Wang, Tian,
  Pan, Min, Feng, Fan, Chen, Wang, Ji, Li, Wang, and Wen}{Zhao
  et~al\mbox{.}}{2020}]%
        {recbole}
\bibfield{author}{\bibinfo{person}{Wayne~Xin Zhao}, \bibinfo{person}{Shanlei
  Mu}, \bibinfo{person}{Yupeng Hou}, \bibinfo{person}{Zihan Lin},
  \bibinfo{person}{Kaiyuan Li}, \bibinfo{person}{Yushuo Chen},
  \bibinfo{person}{Yujie Lu}, \bibinfo{person}{Hui Wang},
  \bibinfo{person}{Changxin Tian}, \bibinfo{person}{Xingyu Pan},
  \bibinfo{person}{Yingqian Min}, \bibinfo{person}{Zhichao Feng},
  \bibinfo{person}{Xinyan Fan}, \bibinfo{person}{Xu Chen},
  \bibinfo{person}{Pengfei Wang}, \bibinfo{person}{Wendi Ji},
  \bibinfo{person}{Yaliang Li}, \bibinfo{person}{Xiaoling Wang}, {and}
  \bibinfo{person}{Ji-Rong Wen}.} \bibinfo{year}{2020}\natexlab{}.
\newblock \showarticletitle{RecBole: Towards a Unified, Comprehensive and
  Efficient Framework for Recommendation Algorithms}.
\newblock \bibinfo{journal}{\emph{arXiv preprint arXiv:2011.01731}}
  (\bibinfo{year}{2020}).
\newblock


\end{thebibliography}

\end{document}